\documentclass[%
 aip,
 pop,
 amsmath,amssymb,
 reprint,%
]{revtex4-1}

\usepackage{graphicx}% Include figure files
\usepackage{dcolumn}% Align table columns on decimal point
\usepackage{bm}% bold math
\usepackage[utf8]{inputenc}
\usepackage[T1]{fontenc}
\usepackage{mathptmx}
\usepackage{etoolbox}
\usepackage{amsmath}
\usepackage[usenames,dvipsnames]{color} % Pureley needed for my TODO command
\usepackage[normalem]{ulem}

\newcommand{\pgrad}{\nabla_\perp}
\newcommand{\BB}{\vec B}
\newcommand{\const}{\mathrm{const}}
\newcommand{\nfp}{\mathcal{N}}

\makeatletter
\def\@email#1#2{%
 \endgroup
 \patchcmd{\titleblock@produce}
  {\frontmatter@RRAPformat}
  {\frontmatter@RRAPformat{\produce@RRAP{*#1\href{mailto:#2}{#2}}}\frontmatter@RRAPformat}
  {}{}
}%
\makeatother
\begin{document}

\preprint{AIP/123-QED}

\title[Compact QS Grad-Shafranov model]{A Grad-Shafranov model for compact quasisymmetric stellarators}
% Force line breaks with \\
\author{N. Nikulsin}
\email{nikita.nikulsin@ipp.mpg.de}
\affiliation{Department of Astrophysical Sciences, Princeton University, Princeton, NJ 08544, USA}
\affiliation{Stellarator Theory Department, Max Plank Institute for Plasma Physics, 17491 Greifswald, Germany}
\author{W. Sengupta}%
\author{S. Buller}
\author{A. Bhattacharjee}
\affiliation{Department of Astrophysical Sciences, Princeton University, Princeton, NJ 08544, USA}

\date{\today}% It is always \today, today,

\begin{abstract}
A Grad-Shafranov equation (GSE) valid for compact quasisymmetric stellarators is derived by an asymptotic expansion around a vacuum field carried to first order. We obtain an equation for the existence of flux surfaces leading up to the GSE. The flux surface label must simultaneously satisfy the existence equation and the GSE, which generally leads to an overdetermined problem. We show how the overdetermined problem can be resolved within our model for a class of hybrid devices similar to that studied by Henneberg and Plunk (S. Henneberg and G. Plunk, PRR 2024). We are also able to solve the existence equation for flux surfaces analytically in the most general case by introducing a special coordinate system. This will enable us to carry out an optimization seeking to minimize the error in our GSE while obeying the flux surface existence equation, which will allow us to find solutions outside the class of hybrid devices. This will allow for a coarse-grained approximate search in the space of quasisymmetric equilibria that should be faster than a conventional stellarator optimization. Nevertheless, it would still be necessary to fine-tune the approximate solutions using conventional tools to obtain a more precise optimized equilibrium.
\end{abstract}

\maketitle

\section{Introduction}

Compact magnetic fusion devices have attracted some interest due to their potential to significantly reduce the cost of a future fusion power plant \cite{sykes2017compact}. While most studies of compact devices have focused on spherical tokamaks \cite{morris2012mast,kaye2019nstx,anand2023modelling}, compact stellarators of the quasi-axisymmetric (QA) type, such as NCSX have also been considered \cite{zarnstorff2001physics}. Recently, a QA device that is a hybrid of a tokamak and stellarator has been proposed \cite{henneberg2024compact}. Such a device would be axisymmetric on the outboard side but have helical corrugations, which would provide a source of external rotational transform on the inboard side. These corrugations are produced by a set of "banana coils," which are also located on the inboard side. Aside from the banana coils, the only other type of coils that are necessary are tokamak-like planar toroidal field coils. Thus, the complexity of coils necessary for constructing such a device is greatly reduced compared to contemporary modular coil stellarators.

It should be noted that the concept of a compact stellarator with helical corrugations on the inboard side and an axisymmetric outboard side has been considered by Moroz already in 1998 \cite{moroz1998helical}, long before Ref \onlinecite{henneberg2024compact}. However, the novelty of Ref \onlinecite{henneberg2024compact} was in showing that such a stellarator can be optimized for quasi-axisymmetry. Quasisymmetry, of which quasi-axisymmetry is a special case, is one of two main approaches to confining particle orbits inside a stellarator \cite{helander2014theory,boozer1998what}. The simplest way to define quasisymmetry is as a stellarator configuration where the magnetic field satisfies $|\BB| = B(\psi, M\theta-N\phi)$, where $M,N$ are arbitrary integers and $(\psi,\theta,\phi)$ is a straight field line coordinate system \cite{helander2014theory}. The case when $N=0$ corresponds to quasi-axisymmetry. Thus, a quasisymmetric magnetic field has a magnitude that depends only on two coordinates, but the full magnetic field vector can still depend on all three coordinates.

A significant complication that arises when designing quasisymmetric stellarators is that the MHD equilibrium equations (with isotropic pressure) already comprise a closed system of equations, and imposing the quasisymmetry condition on $|\BB|$ appears to overdetermine the system, requiring a careful optimization procedure to find a satisfactory configuration \cite{bader2019stellarator}. Within the context of the near-axis expansion \cite{garren1991existence,garren1991magnetic,landreman2018direct,jorge2020construction}, which consists of Taylor-expanding all quantities of interest in the radial coordinate and balancing the coefficients of the MHD equilibrium equations at each order of the expansion, resulting in a system of ordinary differential and algebraic equations, the overconstraining problem is absent at first order \cite{garren1991existence}. At second order, the problem can be dealt with by not imposing quasisymmetry directly, but instead optimizing for it, which is much faster than traditional optimization due to the smaller parameter space \cite{landreman2022mapping}. At third order and beyond, the system of equations is severely overdetermined, and there is evidence that the Taylor series itself begins to diverge \cite{rodriguez2022quasisymmetry}.

The near-axis model, however, cannot be applied to low aspect ratio devices like the Henneberg-Plunk stellarator described in Ref \onlinecite{henneberg2024compact}. In this paper, we will apply a Grad-Shafranov model\footnote{The model consists of two equations for the flux surface label: a GSE, which is a second-order elliptic PDE involving pressure and current, and a flux surface existence condition.} simpler versions of which were considered in Refs \onlinecite{sengupta2024asymptotic,nikulsin2024an}, to study a Henneberg-Plunk-like device. This model is derived in the spirit of the near-axis model, being a simplification of the underlying equations that can aid conceptual understanding or be used as an initial guess for a stellarator optimization. It, however, cannot replace numerical solutions from equilibrium codes like VMEC\cite{hirshman1983steepest} and DESC\cite{dudt2020desc} when high fidelity results are needed.

In previous papers \cite{sengupta2024asymptotic,nikulsin2024an}, the Grad-Shafranov model was derived via an expansion in the inverse aspect ratio. However, a more general expansion in the deviation of the total magnetic field from the coil-generated field is also possible. Indeed, in Ref \onlinecite{nikulsin2024an}, we show that, under the assumptions made in that paper, such an expansion is equivalent to a large-aspect-ratio expansion. It should be noted that a GSE for quasisymmetric stellarators was derived without any expansions or approximations in Ref \onlinecite{burby2020some}. However, that paper does not address the overdetermination problem or attempt to solve the equation. In this paper, we will consider the overdetermination problem head-on and propose an approach to resolving it for a particular kind of compact quasisymmetric device.

The rest of this paper is organized as follows. In section \ref{sec:deriv}, we will derive the model under a slightly different set of assumptions than in Ref \onlinecite{nikulsin2024an}, which allow for compact geometries. Then, in section \ref{sec:HPstell}, we will apply the newly derived model to a Henneberg-Plunk-like stellarator to demonstrate its utility. The approximate solution obtained in section \ref{sec:HPstell} will then be refined via conventional stellarator optimization in section \ref{sec:refinement}. Finally, we conclude in section \ref{sec:concl} by discussing further work that could allow this model to be applied to a broader class of compact stellarators. In addition, in appendix \ref{sec:infgen} we show that the infinitesimal generator of quasisymmetry, as defined in Ref. \onlinecite{rodriguez2020necessary}, has a particularly simple form when written in terms of quantities that are introduced in the present model.

\section{Derivation}\label{sec:deriv}

In Ref \onlinecite{nikulsin2024an}, we expanded the magnetic field around a vacuum magnetic field $\nabla\chi$ ($\chi$ is the magnetic scalar potential):
\begin{equation}
    \BB = \left(1 + \frac{B_1}{B_v}\right)\nabla\chi + \nabla A\times\nabla\chi + O(\epsilon^2),\label{eq:Bexp}
\end{equation}
under the assumption that $B_v = |\nabla\chi| = O(1)$, $B_1 = O(\epsilon)$, where $\epsilon = \max|\BB - \nabla\chi|/B_v \ll 1$ and $A = O(\epsilon)$ is a stream function needed to represent perturbations perpendicular to $\nabla\chi$. As is usually the case in reduced MHD models, we also order the derivative along the vacuum field as $\nabla\chi\cdot\nabla = O(\epsilon)$, whereas $\pgrad = \nabla - B_v^{-1}\nabla\chi\cdot\nabla = O(1)$. As discussed in Ref \onlinecite{nikulsin2024an}, this last ordering can be justified by considering the length scales in each direction. Note that $\chi$ is a multivalued function. As will be seen later, in equation \eqref{eq:chi}, when written in cylindrical coordinates, the multivaluedness enters via the secular term $F_0\phi$, where $F_0 = \const$ and $\phi$ is the toroidal angle.

The current density can be written in two different ways, one of which is obtained from the definition of the current density and the other from the MHD equilibrium equation:
\begin{subequations}
    \begin{gather}
        \vec j = \frac{\BB\times\nabla p}{B^2} + j_\parallel\frac{\BB}{B} = \frac{\nabla\chi\times\nabla p}{B_v^2} - \frac{\BB}{\mu_0}\Delta^* A + O(\epsilon^2),\label{eq:curr1}\\
        \vec j = \frac{1}{\mu_0}\nabla\times\BB = \frac{1}{\mu_0}\nabla\left(\frac{B_1}{B_v}\right)\times\nabla\chi - \frac{\nabla\chi}{\mu_0}\nabla^2 A + O(\epsilon^2),
    \end{gather}\label{eq:curr}
\end{subequations}
where $\Delta^* = B_v^{-2}\nabla\cdot(B_v^2\pgrad) = O(1)$. As can be seen, the leading order current is $O(\epsilon)$, which is consistent with $\nabla A\times\nabla\chi$ being the leading order term in the plasma-current-generated part of the magnetic field.

As we showed in Ref \onlinecite{nikulsin2024an}, in order for the perpendicular components of both equations \eqref{eq:curr} to match, either the aspect ratio must be large, or we must have $B_1 = 0$ and $p = O(\epsilon^2)$. We have already considered the first case in Ref \onlinecite{nikulsin2024an}; we will now study the second case, which allows for compact geometries. Thus, expression \eqref{eq:Bexp} becomes
\begin{equation}
    \BB =\nabla\chi + \nabla A\times\nabla\chi,\label{eq:Bfield}
\end{equation}
which we will use throughout the rest of this paper.

\subsection{Obtaining the pre-Grad-Shafranov equation}

Taking the divergence of equation \eqref{eq:curr1}, we obtain the main equation, which can be transformed to a Grad-Shafranov equation in the case of quasisymmetric stellarators:
\begin{equation}
    \begin{aligned}
        &\BB\cdot\nabla\left(\frac{j_\parallel}{B}\right) + \nabla\cdot\left(\frac{\BB\times\nabla p}{B^2}\right) \\
        &= \frac{-1}{\mu_0}\BB\cdot\nabla\Delta^* A + \nabla\left(\frac{1}{B_v^2}\right)\cdot(\nabla\chi\times\nabla p) + O(\epsilon^3) = 0.
    \end{aligned}\label{eq:divj}
\end{equation}
To close the system, we also need an equation for $p$, which is simply $\BB\cdot\nabla p = 0$.

We can now impose the two-term quasisymmetry condition, which is equivalent to demanding that $|\BB| = B(\psi, M\theta - N\phi)$: $(\BB\times\nabla\psi)\cdot\nabla B = F(\psi)\BB\cdot\nabla B$, where $\psi$ is the toroidal flux \cite{helander2014theory}. Under our ordering, this condition can be written in two equivalent ways:
\begin{subequations}
    \begin{gather}
        (\nabla\chi\times\nabla\Psi)\cdot\nabla B_v = \BB\cdot\nabla B_v,\label{eq:QSA}\\
		(\nabla B_v\times\nabla\chi)\cdot\nabla(\Psi + A) = \nabla\chi\cdot\nabla B_v,\label{eq:QSB}
    \end{gather}
\end{subequations}
where $\Psi = \int d\psi/F(\psi)$ is defined to absorb the $F(\psi)$ factor. Note that \cite{helander2014theory} $F(\psi)$ is $O(\epsilon^{-1})$, so $\Psi = O(\epsilon)$. Using \eqref{eq:QSA}, the second term in the RHS of equation \eqref{eq:divj} can be rewritten as $(dp/d\Psi)(\nabla\chi\times\nabla\Psi)\cdot\nabla B_v^{-2} = (dp/d\Psi)\BB\cdot\nabla B_v^{-2}$, which results in both terms in the equation having $\BB\cdot\nabla$ acting on something. We then remove the $\BB\cdot\nabla$ operator, obtaining a Grad-Shafranov-like equation:
\begin{equation}
    \Delta^* A - \frac{\mu_0}{B_v^2}\frac{dp}{d\Psi} = H(\Psi).\label{eq:preGS}
\end{equation}

\subsection{A coordinate system for near-vacuum quasisymmetry}

We introduce a new coordinate system, $(B_v,s,\chi)$, where $B_v$ and $\chi$ are as defined above and $s$ is an arbitrary third coordinate. One can choose $s$ to be any quantity as long as the Jacobian is nonsingular in the region of interest. This is essentially a generalization of the cylindrical $(R,z,\phi)$ coordinates: recall that in the tokamak limit, $B_v = F_0/R$ and $\chi = F_0\phi$, thus $B_v$ is the radial coordinate and $\chi$ is the toroidal coordinate. In principle, $s$ has no simple interpretation, due to the freedom one has in choosing a third coordinate, however, it can be chosen to be a vertical-like coordinate. Using these coordinates, we can rewrite equation \eqref{eq:QSB} as
\begin{equation}
    \frac{1}{\sqrt{g}}\frac{\partial}{\partial s}(\Psi + A) = -g^{B_v\chi},
\end{equation}
where $g$ is the metric tensor. Integrating over $s$, we obtain an algebraic relation between $A$ and $\Psi$:
\begin{equation}
    \begin{gathered}
        \Psi + A = -I + a(B_v,\chi), \\
        I = \int g^{B_v\chi}\sqrt{g}ds = \int\frac{\nabla\chi\cdot\nabla B_v}{(\nabla B_v\times\nabla s)\cdot\nabla\chi}ds,
    \end{gathered}\label{eq:QSC}
\end{equation}
where $a(B_v,\chi)$ is an arbitrary function that appears due to the integration. 

Now consider the equation $\BB\cdot\nabla p = 0$. Since $p = p(\Psi)$, using equations \eqref{eq:Bfield} and \eqref{eq:QSC}, we have
\begin{equation}
    \BB\cdot\nabla\Psi = B_v^2\frac{\partial\Psi}{\partial\chi} + g^{s\chi}\frac{\partial\Psi}{\partial s} + \frac{1}{\sqrt{g}}\frac{\partial\Psi}{\partial s}\left(\frac{\partial I}{\partial B_v} - \frac{\partial a}{\partial B_v}\right) = 0.\label{eq:B.gps}
\end{equation}
This equation can be further simplified by considering the fact that $\chi$ must satisfy the Laplace equation. Using the expression for divergence in general non-orthogonal coordinates, one has
\begin{equation}
    \begin{aligned}
        \sqrt{g}\nabla^2\chi &= \frac{\partial}{\partial q^k}\left[\sqrt{g}(\nabla\chi)^k\right] \\
        &= \frac{\partial}{\partial B_v}(g^{B_v\chi}\sqrt{g}) + \frac{\partial}{\partial s}(g^{s\chi}\sqrt{g}) + B_v^2\frac{\partial}{\partial\chi}\sqrt{g} = 0.
    \end{aligned}
\end{equation}
Integrating this equation over $s$, we obtain
\begin{equation}
    \frac{\partial I}{\partial B_v} = -g^{s\chi}\sqrt{g} - B_v^2\frac{\partial}{\partial\chi}\int\sqrt{g}ds + \widetilde{a}(B_v,\chi).\label{eq:dIdBv}
\end{equation}
Substituting this into equation \eqref{eq:B.gps} yields:
\begin{equation}
    B_v^2\frac{\partial\Psi}{\partial\chi} - \frac{1}{\sqrt{g}}\frac{\partial\Psi}{\partial s}\left(B_v^2\frac{\partial}{\partial\chi}\int\sqrt{g}ds + \frac{\partial a}{\partial B_v}\right) = 0,\label{eq:B.gpsM}
\end{equation}
where $\widetilde{a}$ has been absorbed into $\partial a/\partial B_v$. This equation can be solved analytically using the method of characteristics. The corresponding Lagrange-Charpit equation is
\begin{equation}
    \begin{aligned}
        \frac{ds}{d\chi} &= \frac{-1}{B_v^2\sqrt{g}}\left(B_v^2\frac{\partial}{\partial\chi}\int\sqrt{g}ds + \frac{\partial a}{\partial B_v}\right) \\
        &= -\left(\frac{\partial U}{\partial s}\right)^{-1}\frac{\partial U}{\partial\chi} - \frac{1}{B_v^2}\left(\frac{\partial U}{\partial s}\right)^{-1}\frac{\partial a}{\partial B_v},
    \end{aligned}
\end{equation}
where $U = \int\sqrt{g}ds$. Multiplying through by $\partial U/\partial s$, the above can be rewritten as
\begin{equation}
    \frac{dU}{d\chi} = \frac{\partial U}{\partial s}\frac{ds}{d\chi} + \frac{\partial U}{\partial\chi} = -\frac{1}{B_v^2}\frac{\partial a}{\partial B_v},
\end{equation}
since $dB_v/d\chi = 0$, as equation \eqref{eq:B.gpsM} does not have a $\partial\Psi/\partial B_v$ term. Thus, the characteristics are
\begin{equation}
    \begin{gathered}
        C_1 = B_v, \\
        C_2 = U + \frac{1}{B_v^2}\int\frac{\partial a}{\partial B_v}d\chi = \int\sqrt{g}ds + \frac{1}{B_v^2}\int\frac{\partial a}{\partial B_v}d\chi,
    \end{gathered}\label{eq:char}
\end{equation}
and the solution of equation \eqref{eq:B.gpsM} is $\Psi = \Psi(C_1, C_2)$.

Finally, note that, since this model relies on quasisymmetry being satisfied at order $\epsilon$, it needs flux surfaces to be nested in order to be valid, with the only exception being island chains that resonate with the helicity of the quasisymmetry\cite{rodriguez2021islands,rodriguez2022quasisymmetry}. While quasisymmetry can be broken at order $\epsilon^2$ and the nested flux surfaces can break at that order, the present model cannot say anything about those effects as it is only an $O(\epsilon)$ model.

\subsection{The final Grad-Shafranov equation}

To obtain the Grad-Shafranov equation, we insert relation \eqref{eq:QSC} into the pre-Grad-Shafranov equation \eqref{eq:preGS}, yielding
\begin{equation}
    \Delta^*\Psi + \Delta^*(I - a) = -\frac{\mu_0}{B_v^2}\frac{dp}{d\Psi} - H(\Psi).\label{eq:GS}
\end{equation}

Note that $\Psi$ is overconstrained: it must simultaneously satisfy both equations \eqref{eq:GS} and \eqref{eq:B.gpsM}. Formulated another way, $\Psi$ must be a function of only the characteristics $C_1$ and $C_2$, as given by the expressions \eqref{eq:char}, while the terms on the LHS of equation \eqref{eq:GS} can, in general, be functions of all three variables. In the next section, we will consider a special case where the overconstraining problem is resolved. Solving the problem in general would require implementing a numerical solver that constructs a function basis in $C_1,C_2$ space and then looking for functions that minimize the error in equation \eqref{eq:GS}. We will return to this task in future work.

Finally, we note that the $\Delta^*$ operator can be written in the $(B_v,s,\chi)$ coordinates as
\begin{equation}
    \begin{aligned}
        \Delta^* &= \frac{g^{B_v B_v}}{B_v^2}\frac{\partial}{\partial B_v}\left(B_v^2\frac{\partial}{\partial B_v}\right) + \frac{2g^{B_v s}}{B_v}\frac{\partial}{\partial B_v}\left(B_v\frac{\partial}{\partial s}\right) \\
        &+ g^{ss}\frac{\partial^2}{\partial s^2} + \Delta B_v\frac{\partial}{\partial B_v} + \Delta s\frac{\partial}{\partial s} + O(\epsilon).
    \end{aligned}\label{eq:GSO}
\end{equation}

\section{A hybrid Henneberg-Plunk-like compact QAS: an application}\label{sec:HPstell}

To illustrate the use of the model derived in the previous section, we will consider a device similar to the compact quasiaxisymmetric stellarator configuration presented in Ref \onlinecite{henneberg2024compact}. We do not attempt to model the exact device presented in Ref \onlinecite{henneberg2024compact} but rather study a similar device based on the same design principles, namely, an axisymmetric outboard side and helical corrugations on the inboard side.

\begin{figure}
    \centering
    \includegraphics[width=\linewidth]{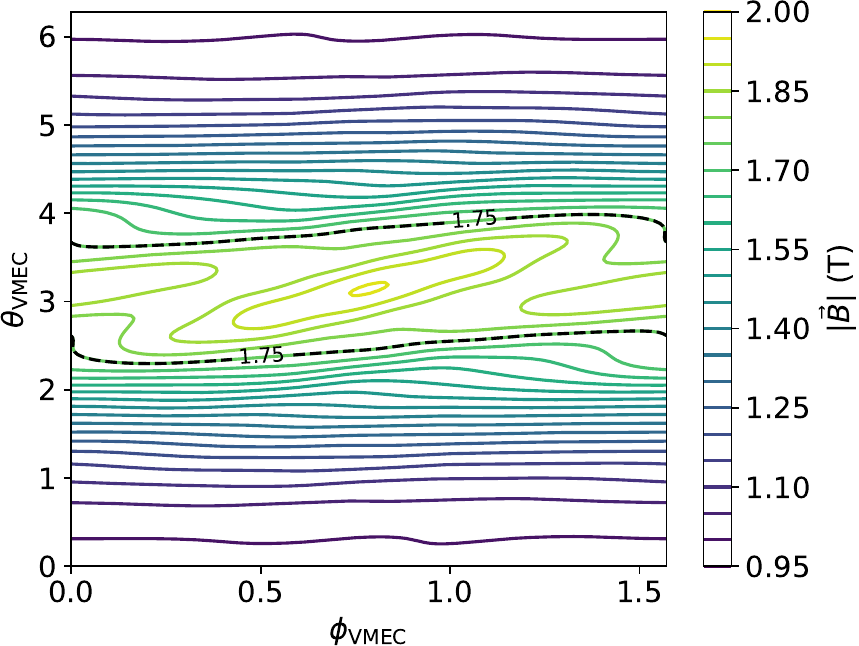}
    \caption{The magnetic field strength on the outermost flux surface of the Henneberg-Plunk stellarator\cite{henneberg2024compact}. The field strength in between the two dashed lines ranges from 1.75~T to 2~T.\\
    Adapted from S. Henneberg and G. Plunk, \emph{Physical Review Research}, Vol. 6, L022052, 2024; licensed under a Creative Commons Attribution (CC BY) license.}
    \label{fig:modB_HP}
\end{figure}

\begin{figure*}
    \centering
    \includegraphics[width=0.18\linewidth]{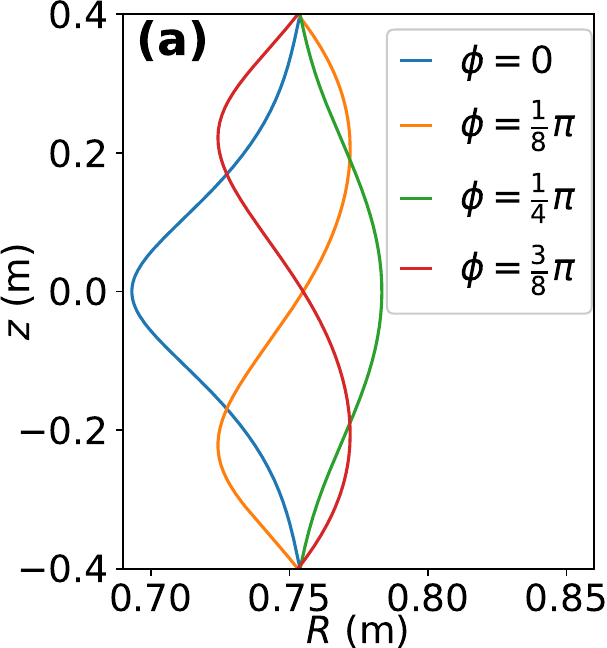} \includegraphics[width=0.265\linewidth]{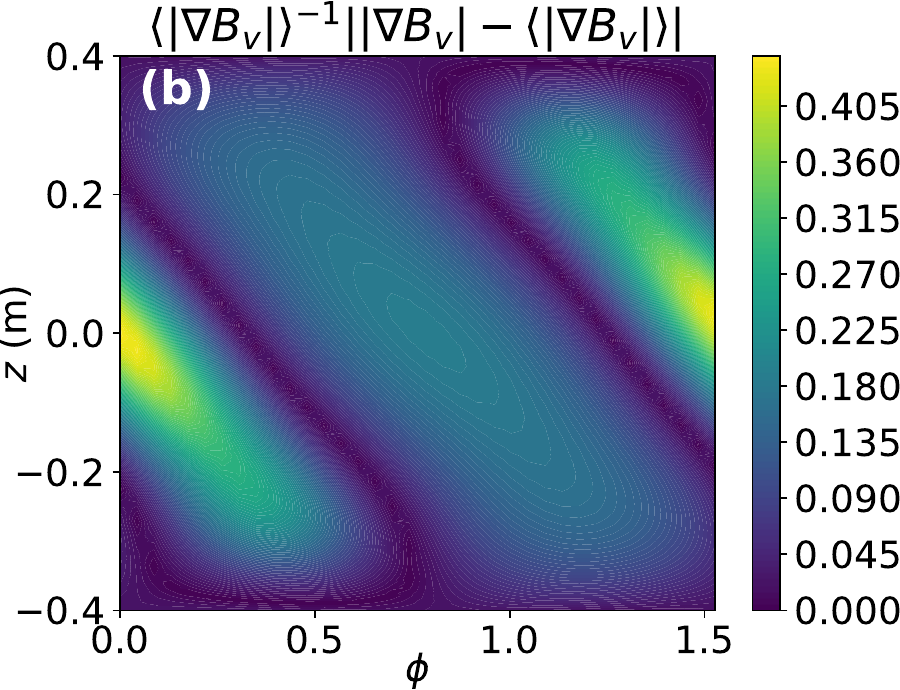} \includegraphics[width=0.265\linewidth]{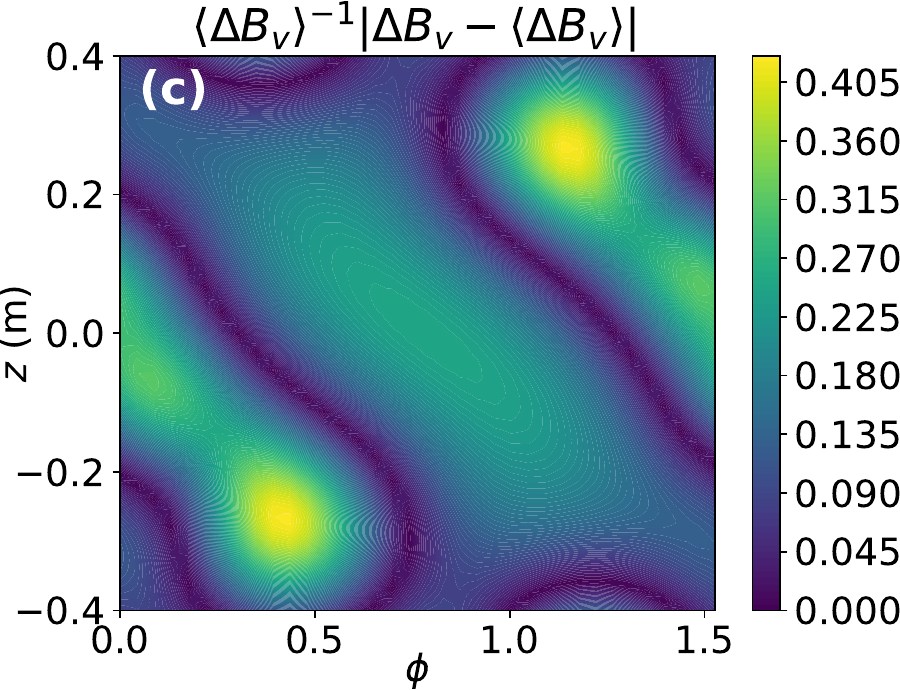} \includegraphics[width=0.265\linewidth]{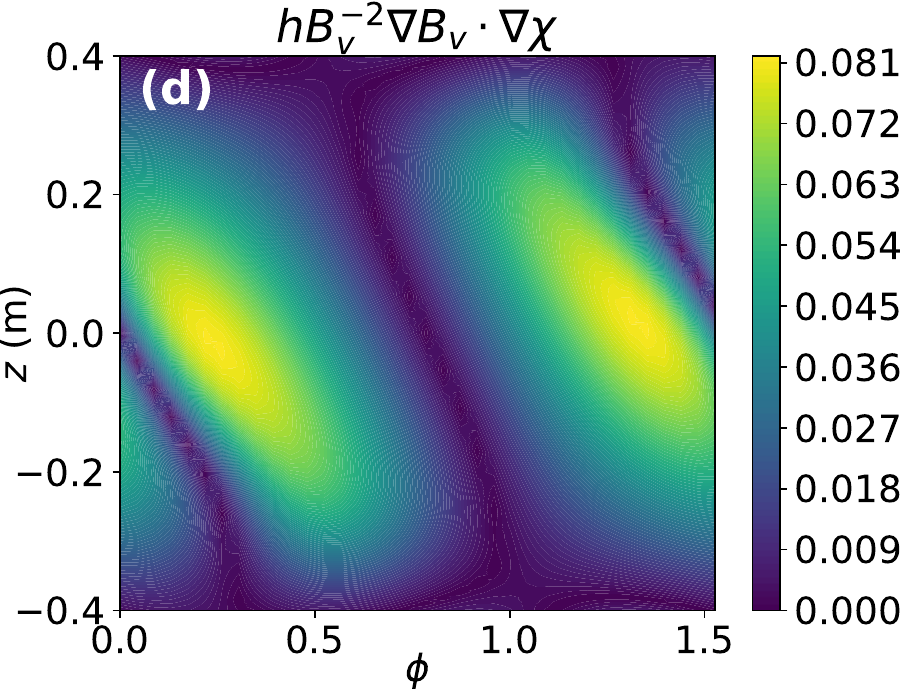}
    \caption{The cross sections of the $B_v = 1.465$~T surface are shown at various values of $\phi$ in (a). In (b), the relative deviation of $|\nabla B_v|$ from its average value on the $B_v = 1.465$~T surface, $\langle|\nabla B_v|\rangle$; the same is shown for $\Delta B_v$ in (c). Finally, (d) shows that the normalized value of $g^{B_v\chi} = \nabla B_v\cdot\nabla\chi$ is small. The quantities in panels (b), (c), and (d) are all shown on the $B_v = 1.465$~T surface, where these quantities are the highest.}
    \label{fig:vacfield}
\end{figure*}

As was discussed in the previous section, one of the main difficulties with quasisymmetry is that the corresponding system of equations is overconstrained. However, as is well known, this overconstraining problem does not arise in the axisymmetric case. Indeed, since the metric tensor of cylindrical coordinates, as well as any other orthogonal coordinate system in Euclidean space that includes the toroidal angle $\phi$ as a coordinate, is independent of $\phi$\footnote{In the language of differential geometry, it can be said that $\vec e_\phi$, the infinitesimal generator of axisymmetry, is a Killing vector.}, one can simply set $\partial/\partial\phi = 0$ and proceed to solve the equations. However, general quasisymmetry does not have this property, and so if one tries to set $\partial/\partial l = 0$, where $l$ is the distance along the direction of quasisymmetry, $l$ will still be present in the equations due to the metric tensor depending on it, resulting in an overconstrained system. With this in mind, we only need to understand how to resolve the overconstraining problem on the inboard side, where the nonaxisymmetric corrugations are present. To do this, we take a closer look at the magnetic field strength on the outermost flux surface of the Henneberg-Plunk device (Figure \ref{fig:modB_HP}), where quasisymmetry is the worst\cite{henneberg2024compact}. Note that outside of the dashed lines, the contours of field strength can be approximated by straight horizontal lines, like in a tokamak. On the other hand, inside the dashed lines, the helical nature of the inboard side manifests via the helically slanted closed contours. This is where the quasisymmetry breaks, as point maxima of magnetic field strength, such as those at the centers of these closed contours, cannot exist in a perfectly quasisymmetric device\cite{helander2014theory}. However, in between the dashed lines, the field strength ranges only from 1.75~T to 2~T, or about 0.24 of the total range seen in the figure. Thus, assuming $\epsilon \sim 0.2$, quasisymmetry can still be satisfied at the lowest order by approximating the field strength in between the dotted lines as constant. Assuming that everything outside of the dotted lines is in the axisymmetric region and that $\chi$ satisfies certain conditions, which we will derive shortly, the overconstraining problem is resolved. As a sidenote, a similar idea of having patches of constant $|\BB|$ on a flux surface was considered in Ref \onlinecite{velasco2024piecewise}, where Velasco et al introduce the concept of a piecewise omnigenous stellarator. Such a field with constant strength can be considered as approximately locally isodynamic \cite{sengupta2021vacuum_NS}. Although exact globally isodynamic configurations can only be axisymmetric or helically symmetric \cite{palumbo1968some,schief2003nested}, local approximate isodynamic constraints can be nonaxisymmetric \cite{sengupta2021vacuum_NS}.

We proceed under the assumption that in the non-axisymmetric region $\partial\Psi/\partial C_2 = O(\epsilon^2)$, i.e. $\Psi$ only depends on $C_1 = B_v$ at the leading order, $O(\epsilon)$, and the $C_2$-dependence only appears at the next order, $O(\epsilon^2)$, and thus $|\BB| = B_v + O(\epsilon^2)$ is nearly constant on each flux surface patch that intersects this region, as seen in Figure \ref{fig:modB_HP}. After setting\footnote{Note that if $\partial^2 a/\partial B_v\partial\chi\neq 0$, then the method of resolving the overconstraining problem used in this paper would not be applicable, as a $\chi$-dependent term would be introduced into equation \eqref{eq:GS_nonax}, whereas the other terms and $\Psi$ itself are only supposed to depend on $B_v$ in the non-axisymmetric region. Aside from that, $a$ is arbitrary and we choose $a=0$ for simplicity.} $a = 0$ and using expression \eqref{eq:GSO}, the Grad-Shafranov equation \eqref{eq:GS} in this region becomes
\begin{equation}
    \frac{g^{B_v B_v}}{B_v^2}\frac{d}{dB_v}\left(B_v^2\frac{d\Psi}{dB_v}\right) + \Delta B_v\frac{d\Psi}{dB_v} = -\frac{\mu_0}{B_v^2}\frac{dp}{d\Psi} - H(\Psi).\label{eq:GS_nonax}
\end{equation}
The RHS will be a function of only $B_v$, thus the LHS must be as well. Note that axisymmetry is broken in the equation via the non-axisymmetric nature of $B_v$. The most straightforward way of achieving this is to demand that $g^{B_v B_v} \equiv |\nabla B_v|^2 = f_1(B_v) + O(\epsilon)$ and $\Delta B_v = f_2(B_v) + O(\epsilon)$, where $f_1$ and $f_2$ are arbitrary functions. In addition, we show that $I = O(\epsilon^2)$, which allows us to drop the $\Delta^* I$ term in equation \eqref{eq:GS_nonax}, is obtained without us needing to impose a stringent constraint on it (Figure \ref{fig:vacfield}~d). Instead, we just impose a penalty on values of $g^{B_v\chi}$ above a certain threshold to prevent the optimization from drifting towards large values of $I$. However, the $g^{B_v\chi}$ in the final result is significantly below the threshold. Note that all of the constraints stated above only affect the vacuum scalar potential $\chi$. Thus, once we find a vacuum magnetic field that satisfies these constraints to a sufficient degree, we can proceed to solve the Grad-Shafranov equations in the nonaxisymmetric and axisymmetric regions using standard methods.

To find an appropriate vacuum scalar potential, we must first choose a basis in the space of solutions to the Laplace equation, in which we will look for the scalar potential. We will choose our basis to be the following subset of cylindrical harmonics\cite{morse1953methods}
\begin{equation}
    \begin{aligned}
        \chi &= F_0\phi + \chi_b,\\
        \chi_b &= \sum_{n=1}^N\int_0^{k_\mathrm{max}} K_n(kR)(A_n(k)\cos kz\cos n\nfp\phi \\
        &+ B_n(k)\cos kz\sin n\nfp\phi + C_n(k)\sin kz\cos n\nfp\phi \\
        &+ D_n(k)\sin kz\sin n\nfp\phi)dk,
    \end{aligned}\label{eq:chi}
\end{equation}
where $\nfp$ is the number of field periods, and $(R,z,\phi)$ are the standard cylindrical coordinates. This solution is obtained by solving the Laplace equation by separation of variables; $-k^2$ and $-n^2$ are eigenvalues of the operators $\partial^2/\partial z^2$ and $\partial^2/\partial\phi^2$, respectively. Due to the periodicity requirement in $\phi$, $n$ must be real and have a discrete spectrum, while $k$ has a continuous spectrum and can be either real or imaginary. We chose to have the $R$-dependence be via the modified Bessel function of the second kind $K_n$, which corresponds to $k$ being real, rather than the ordinary Bessel functions $J_n$ and $N_n$ because the "banana coils" of the Henneberg-Plunk device are roughly helical. The magnetic field generated by a helical wire can be shown to depend on $R$ via modified Bessel functions\cite{tominaka2001analytical}. We also excluded the modified Bessel function of the first kind $I_n$ from the basis, as it goes to infinity as $R\to\infty$, whereas we need the field to decay to zero far from the banana coils. The general solution is a linear combination of particular solutions, obtained by integrating over the continuous eigenvalue and summing over the discrete one, with $A_n(k)$, $B_n(k)$, $C_n(k)$ and $D_n(k)$ being the coefficients. Finally, we introduce an additional constraint on the basis, namely that $\partial\chi_b/\partial\phi|_{z=\pm h} = 0$, where $h$ is a constant. The purpose of this constraint is to ensure that there are two $z = \const$ planes where the value of $B_v$ is roughly independent of $\phi$, so that the nonaxisymmetric, $B_v$-dependent part of a flux surface can connect to the axisymmetric part in a small region where $z = \pm h + O(\epsilon)$. Note that, since $B_v^2 = (\partial\chi/\partial R)^2 + (\partial\chi/\partial z)^2 + R^{-2}(\partial\chi/\partial\phi)^2$, there can still be a $\phi$-dependence in $B_v$ via $\partial\chi/\partial z$, which does not have to be independent of $\phi$ at $z = \pm h$ even if $\partial\chi_b/\partial\phi$ is zero, however, in practice this $\phi$-dependence is negligible (Figure \ref{fig:vacfield}~a). After imposing this condition, the eigenvalue $-k^2$, which could previously be continuous, can now only take a discrete set of values. After also imposing stellarator symmetry ($\chi(R,-z,-\phi) = -\chi(R,z,\phi)$), we obtain the basis that we will use:
\begin{equation}
    \begin{aligned}
        \chi_b &= \sum_{n=1}^N\sum_{m=0}^M \Bigg[b_{nm}K_n\left(\frac{2m+1}{2h}\pi R\right)\cos\left(\frac{2m+1}{2h}\pi z\right)\sin n\nfp\phi \\
        &+ c_{nm}K_n\left(\frac{m+1}{h}\pi R\right)\sin\left(\frac{m+1}{h}\pi z\right)\cos n\nfp\phi\Bigg].
    \end{aligned}
\end{equation}

Having decided on a basis, the series is truncated at $N=1$ and $M=2$, and initial guesses for the parameters $F_0$, $b_{nm}$ and $c_{nm}$ are determined by performing a least squares fit of $\nabla\chi$ with $h=0.4~\mathrm{m}$ to the vacuum magnetic field of the Henneberg-Plunk device, as calculated by a Biot-Savart code from the banana coil data. More specifically, we minimize the value of the integral $\int_\mathcal{V} |\nabla\chi - \BB_{BS}|^2dV$, where $\mathcal{V}$ is defined as the subvolume of $0.55~\mathrm{m} \leq R \leq 1.8~\mathrm{m}$, $-0.8~\mathrm{m} \leq z \leq 0.8~\mathrm{m}$ where $|\BB_{BS}|^{-1}h|\nabla\cdot\BB_{BS}| \leq 0.05$, i.e. where the nonphysical divergence in the Biot-Savart field resulting from numerical error is sufficiently small. As this is a linear problem where the unknowns are a set of scalars, the linear algebraic equations for $F_0$, $b_{mn}$, and $c_{mn}$ are trivial to derive and solve.

\begin{figure}[b]
    \centering
    \includegraphics[width=0.485\linewidth]{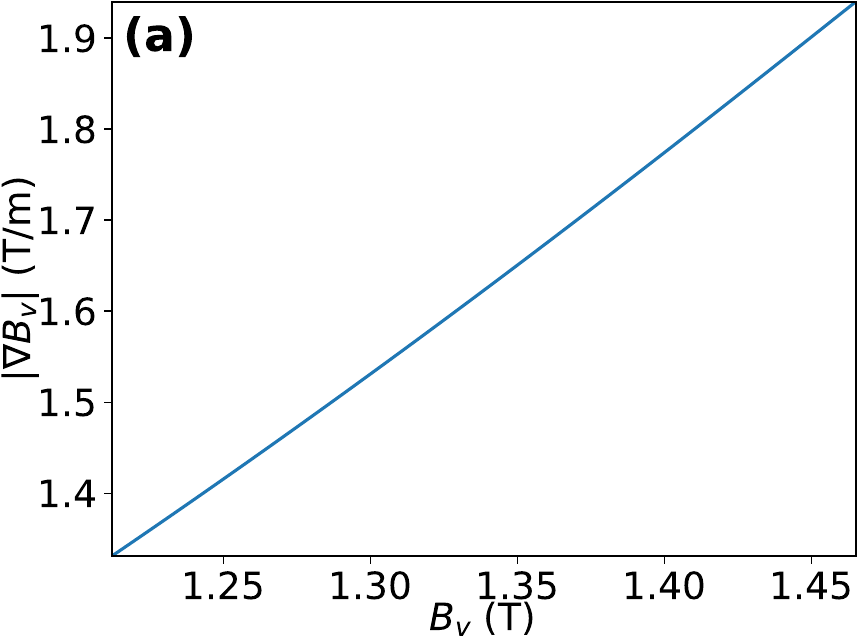}
    \includegraphics[width=0.505\linewidth]{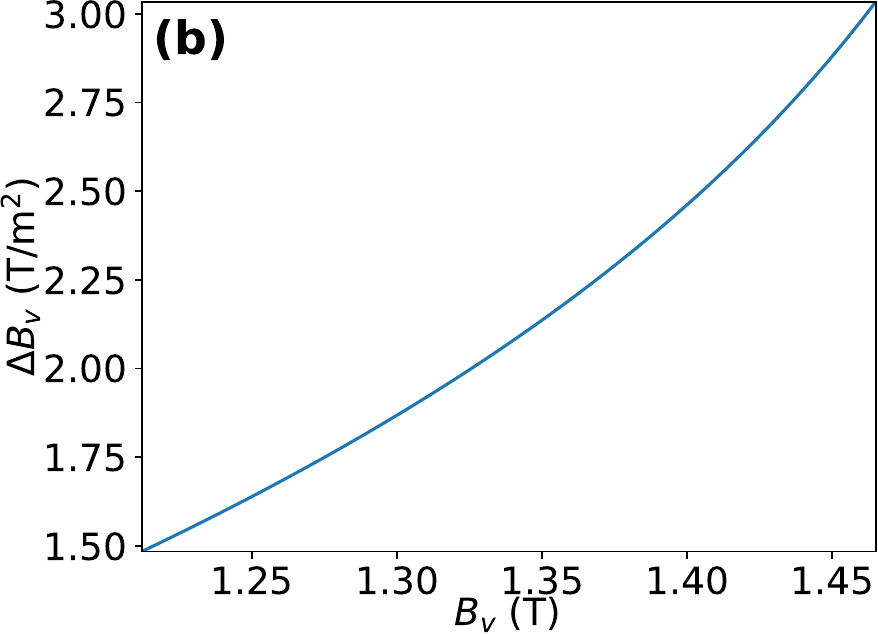}
    \caption{The quantities $|\nabla B_v|$ (a) and $\Delta B_v$ (b), averaged on each $B_v = \const$ surface.}
    \label{fig:Bv_funcs}
\end{figure}

In the next step, we try to minimize the dependence of $g^{B_v B_v}$ and $\Delta B_v$ on variables other than $B_v$ by varying $b_{nm}$ and $c_{nm}$ while holding $F_0$ fixed. We seek to minimize the following objective with a least-squares approach, using the Trust Region Dogleg approach\cite{voglis2004rectangular}, as implemented in the SciPy library\cite{virtanen2020scipy}:
\begin{equation}
    \begin{aligned}
        &\left[\max_{\vec r\in\mathcal{W}} h^2\frac{|(\nabla B_v\times\nabla\chi)\cdot\nabla|\nabla B_v||}{B_v|\nabla B_v\times\nabla\chi|}\right]^2 \\
        + &\left[\max_{\vec r\in\mathcal{W}} h^3\frac{|(\nabla B_v\times\nabla\chi)\cdot\nabla\Delta B_v|}{B_v|\nabla B_v\times\nabla\chi|}\right]^2 \\
        + &\left[R\left(\max_{\vec r\in\mathcal{W}} h\frac{|\nabla\chi\cdot\nabla B_v|}{B_v^2} - 4\right)\right]^2.
    \end{aligned}\label{eq:obj}
\end{equation}
Here, $R(x) = \max\{0,x\}$ is the ramp function and $\mathcal{W}$ is defined as the subvolume of $0.53~\mathrm{m} \leq R \leq 1~\mathrm{m}$, $-0.4~\mathrm{m} \leq z \leq 0.4~\mathrm{m}$ where $B_v \leq 1.4~\mathrm{T}$. This optimization was lightweight enough to be done on a laptop, with the JAX automatic differentiation framework\cite{jax2018github} being used to calculate the derivatives in \eqref{eq:obj}. As shown in Figure \ref{fig:vacfield}, the resulting $|\nabla B_v|$ and $\Delta B_v$ both have a maximum deviation of $\sim 0.4\sim 2\epsilon$ from their respective average values on a $B_v = \const$ surface in the region of interest, which is defined as all points between $z=-0.4$~m and $z=0.4$~m where $B_v\leq 1.465$~T. The average values of $|\nabla B_v|$ and $\Delta B_v$, to be used as coefficients in equation \eqref{eq:GS_nonax}, are shown in Figure \ref{fig:Bv_funcs}. The outermost flux surface will follow the $B_v = 1.465$~T surface in the nonaxisymmetric region. In addition, $hB_v^{-2}\nabla\chi\cdot\nabla B_v \leq 0.08 \sim 2\epsilon^2$, meaning that $I = O(\epsilon^2)$ and can be dropped from equation \eqref{eq:GS}.

As discussed above, the last component in \eqref{eq:obj} is added to prevent the optimization from drifting towards large values of $I$ by penalizing values of the normalized $g^{B_v\chi}$ above 4. However, the final $g^{B_v\chi}$ is far below the threshold. The final set of vacuum field coefficients that we will work with is given in Table \ref{tab:chi_coefs}.

\begin{table}[h]
    \centering
    \caption{The set of vacuum field coefficients (in units of $\mathrm{T\cdot m}$) obtained by optimization.}
    \label{tab:chi_coefs}
    \begin{ruledtabular}   
        \begin{tabular}{c|c|c|c}
            $F_0$ & -1.1031067415616023 & & \\
            $b_{10}$ & 0.035820217604140225 & $c_{10}$ & 1.7606108849811601 \\
            $b_{11}$ & 15.797529271426379 & $c_{11}$ & -75.55668715202704 \\
            $b_{12}$ & -341.8770726097677 & $c_{12}$ & 6833.099190151348
        \end{tabular}
    \end{ruledtabular}
\end{table}

We now proceed to solve the Grad-Shafranov equation numerically, with $p(\Psi) = p_1\Psi$ and $H(\Psi) = H_0 + H_1\Psi$, where $p_1 = 3\cdot 10^6/(2\pi)$~Pa/m, $H_0 = 0.25$ and $H_1 = 0$. We first consider the nonaxisymmetric region, where the Grad-Shafranov equation reduces to the ODE \eqref{eq:GS_nonax}. This equation is solved on the interval $B_v\in[1.212~\mathrm{T},1.465~\mathrm{T}]$, where the outermost flux surface is at $B_v=1.465$~T. Due to the fact that $\chi$ decays exponentially with $R$, $B_v = \const$ surfaces approach $R = \const$ surfaces as $R$ increases. On the $B_v = 1.465$~T surface, the point at $z=0$ moves 9.2~cm as $\phi$ changes from 0 to $\pi/4$ (Figure \ref{fig:vacfield}~a), whereas at $B_v = 1.212$~T the same point only moves by 3.05~cm or $\sim 2\epsilon^2$ of the minor radius. We will neglect this variation and try to patch the axisymmetric and nonaxisymmetric regions at $R=0.91$~m, which is the approximate location of the $B_v=1.212$~T surface, making it the boundary between the two regions.

We impose a boundary condition of $\Psi = 0$. In addition, when solving the ODE \eqref{eq:GS_nonax}, we have one more degree of freedom (either the value of $\Psi$ at $B_v = 1.212$~T or its derivative at either endpoint) that needs to be fixed in order to have a unique solution. We choose $\partial\Psi/\partial B_v = 0.02$~m/T at $B_v = 1.465$~T as an initial guess. We will refine the solution iteratively in tandem with the solution in the axisymmetric region in a way that minimizes the surface current on the boundary between the two regions. The initial solution is then found using the Runge-Kutta 5(4) method\cite{dormand1980a}.

\begin{figure}
    \centering
    \includegraphics[width=0.5\linewidth]{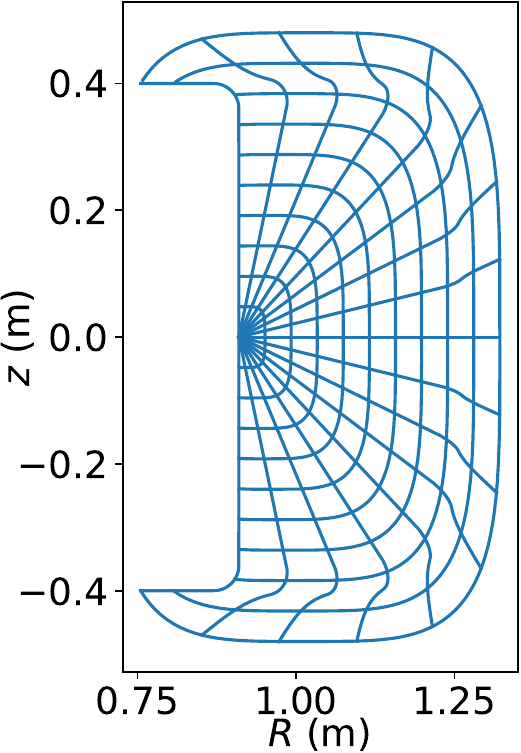}
    \caption{The finite element grid that was used to solve equation \eqref{eq:GS_ax}. The resolution is reduced for clarity.}
    \label{fig:grid}
\end{figure}

In the axisymmetric region, the Grad-Shafranov equation \eqref{eq:GS} reduces to the familiar tokamak Grad-Shafranov equation
\begin{equation}
    \Delta^*\Psi = -\frac{\mu_0 R^2}{F_0^2}\frac{dp}{d\Psi} - H(\Psi),\label{eq:GS_ax}
\end{equation}
with the only difference being that the poloidal flux used in the standard form of the Grad-Shafranov equation has been replaced with the flux function $\Psi$. To solve this PDE, we will use a custom-made finite element solver. The domain on which we look for a solution is bounded by a superellipse given by the equation
\begin{equation}
    \frac{(R-1.015~\mathrm{m})^4}{(0.3065~\mathrm{m})^4} + \frac{z^4}{(0.48~\mathrm{m})^4} = 1,
\end{equation}
with the section $\{(R,z):R<0.91~\mathrm{m}, |z|<0.4~\mathrm{m}\}\setminus\{(R,z):R>0.87~\mathrm{m}, |z|>0.36~\mathrm{m}, (R-0.87~\mathrm{m})^2 + (|z|-0.36~\mathrm{m})^2 > (0.04~\mathrm{m})^2\}$ removed. The second set being subtracted accounts for the corners, which would otherwise have been at $R=0.91$~m, $z=\pm 0.4$~m, being filleted with a radius of 0.04~m to avoid coordinate singularities. The resulting FEM grid is shown in Figure \ref{fig:grid}, with resolution reduced for clarity; the actual resolution was 50 radially and 32 poloidally. Dirichlet boundary conditions are imposed, with $\Psi=0$ on the external boundary and $\Psi$ being matched to the nonaxisymmetric solution on the internal boundary. The axisymmetric solution is then found using the Bubnov-Galerkin method; the basis functions are cubic Bezier splines with $G^1$ continuity at element edges. Finally, the average value of the derivative $\partial\Psi/\partial B_v$ on the line segment $R=0.91$~m, $|z|<0.36$~m (i.e., the part of the boundary between the axisymmetric and nonaxisymmetric regions that is between the two filleted corners) is calculated. The solution of \eqref{eq:GS_nonax} is then recalculated using the SciPy\cite{virtanen2020scipy} implementation of a fourth-order collocation method discussed in Ref \onlinecite{kierzenka2001a}, with this value as a boundary condition for $\partial\Psi/\partial B_v$ at $B_v=1.212$~T, while maintaining the boundary condition of $\Psi=0$ at $B_v=1.465$~T. This new nonaxisymmetric solution is then used to update the boundary conditions for the axisymmetric solution on the internal boundary, followed by recalculating the axisymmetric solution. This process is repeated until the value of $\partial\Psi/\partial B_v$ at $B_v=1.212$~T converges to within 1\%. Note that while this procedure minimizes the surface current on the internal boundary, it does not completely eliminate it, as the $\partial\Psi/\partial B_v$ on the axisymmetric side can vary with $z$, and the corresponding derivative on the nonaxisymmetric side is only matched to its average value, while local discontinuities still remain.

\begin{figure}[b]
    \centering
    \includegraphics[width=0.75\linewidth]{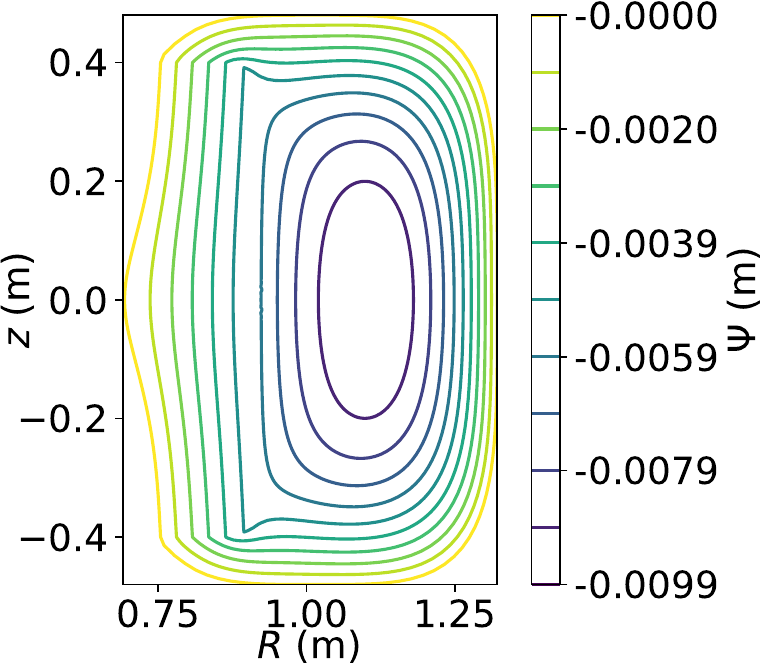}
    \caption{The flux surfaces of the resulting approximate solution on the $\phi=0$ plane, color coded by the value of $\Psi$ on each surface.}
    \label{fig:flux_surfaces}
\end{figure}

The flux surfaces of the resulting solution are shown in Figure \ref{fig:flux_surfaces}. Some of the flux surfaces have sharp corners, which is an artifact of the imposition of a sharp boundary between the axisymmetric and nonaxisymmetric regions. In reality, a boundary layer would form between the two regions, smoothing the corners and alleviating the surface current issue discussed above. However, we will not attempt to find the boundary layer correction to the solution shown in Figure \ref{fig:flux_surfaces}, and instead will just use it as an initial guess for a stellarator optimization, which we will present in the next section.

\section{Refinement via conventional stellarator optimization}\label{sec:refinement}

In terms of quasisymmetry, the solution presented in the previous section has a quasisymmetry error that is several times higher than the configuration of Henneberg-Plunk \cite{henneberg2024compact}. Here, the error on a flux surface $s$ is defined as
\begin{equation}
    \sqrt{\sum_{\substack{m \\ n\neq 0}}\widehat{B}_{n,m}(s)^2\Big/\widehat{B}_{0,0}(s)^2}
\end{equation}
where $\widehat{B}_{n,m}(s)$ are the Fourier modes of $|\BB|$ on the flux surface.

However, by using the solution as the initial condition in a conventional stellarator optimization, we are able to bring the average value of the error, as defined above, down to 0.0158, which is comparable to Henneberg-Plunk (average error 0.0104).

\begin{figure}[b]
\includegraphics[width=\linewidth]{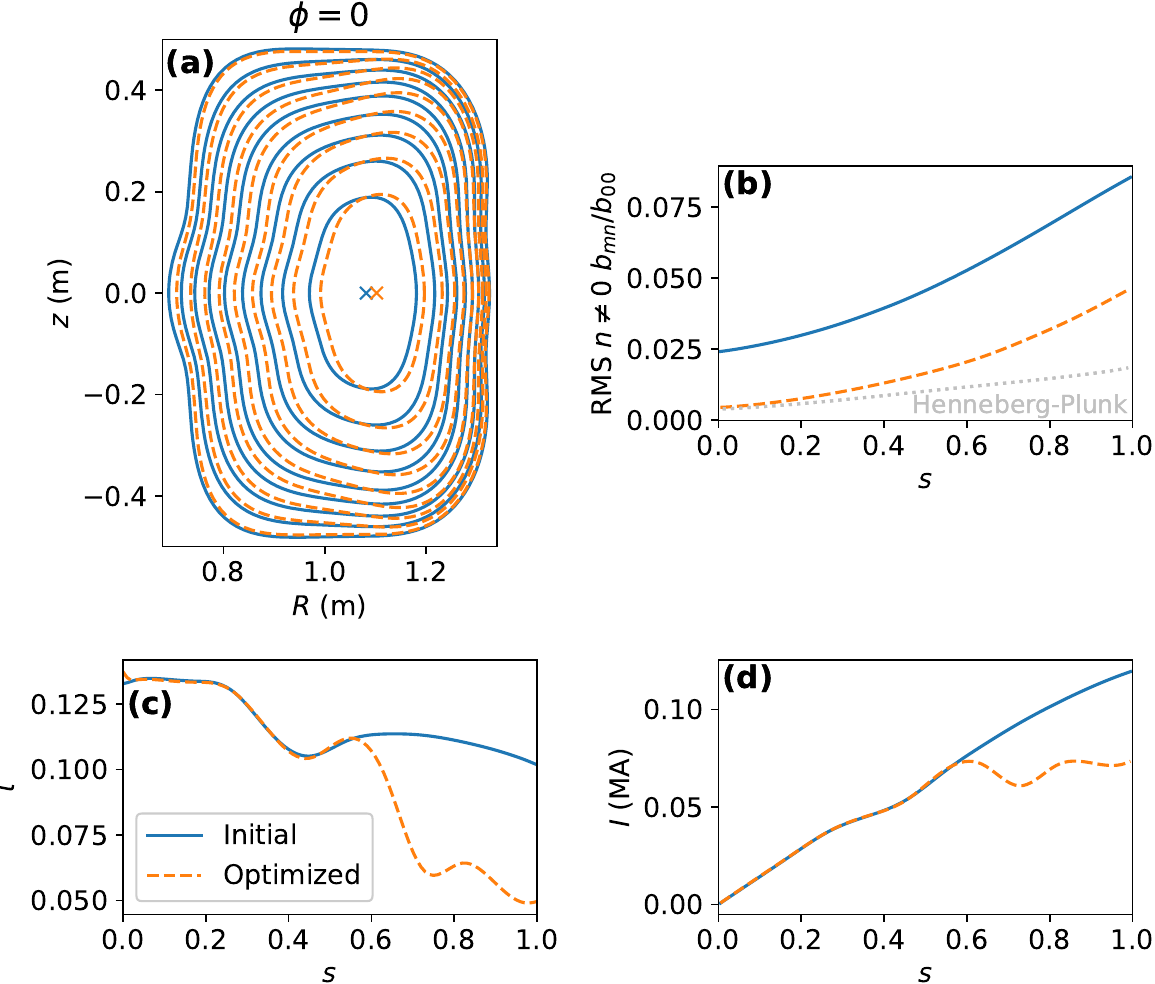}
\caption{\label{fig:optVsIni} Comparison of (a) flux surface shapes of the initial and optimized configuration and (b) quasisymmetry error. Also shown are the (c) $\iota$ and (d) enclosed current $I$. The initial configuration is obtained by passing the boundary and current from the solution in section \ref{sec:HPstell} to VMEC. Both the initial and optimized configurations are calculated with a toroidal resolution of $n=0,N,2N$ and a poloidal resolution of $m=0,1,...,9$.}
\end{figure}

Specifically, we truncate the Fourier representation of the initial condition to $n=0,N,2N$ toroidally and $m=0,1,...,9$ poloidally, where $N=4$ is the number of field periods, and represent the enclosed current $I(s)$ as a cubic spline sampled at 12 points, to reduce the number of degrees of freedom. We then use SIMSOPT\cite{landreman2021simsopt} to minimize the following objective:
\begin{equation}
\begin{aligned}
    \sum_{m,n \neq 0} \frac{b_{mn}(s=1)^2}{b_{00}(s=1)^2} &+ 50\ \max_i I(s_i)^2 + (\bar\iota - \iota_*)^2 \\
    &+ 50\ R\left(0.05 - \min_i\iota(s_i)\right)^2,
\end{aligned}\label{eq:objective}
\end{equation}
where the first term represents the symmetry-breaking part of the Boozer spectrum at the boundary, and the second term minimizes the total toroidal current inside equispaced discrete radii $s_i = i/(N-1), i=0,\dots N-1$, with $I(s_i)$ being the enclosed current in megaamperes. The third term keeps the value of $\bar\iota$, which is the rotational transform averaged over all flux surfaces, except for the axis and six flux surfaces closest to it, close to $\iota_*$, which we take to be the mean $\iota$ of the initial condition, again with the axis and the six nearest flux surfaces excluded. The exclusion of these flux surfaces is necessary to avoid inaccuracies due to the coordinate singularity on the VMEC axis\cite{panici2023the}. Finally, the last term penalizes the $\iota$ dipping below 0.05 at any point, with $R$ being the ramp function, as defined previously. Note that there is nothing in the minimization objective \eqref{eq:objective} that forces the configuration to be compact or tokamak-like: these features are purely a result of starting the optimization from a suitable initial condition. A comparison of the initial and optimized configurations is shown in Figure \ref{fig:optVsIni}.

\section{Conclusion}\label{sec:concl}

We have derived an asymptotic model that is valid for compact quasisymmetric stellarators with $\beta = O(\epsilon^2)$, under the assumption that the vacuum magnetic field is dominant ($\epsilon\ll 1$). However, non-vacuum effects can still contribute ($\nabla\chi\cdot\nabla = O(\epsilon)$). Just like in the large aspect ratio high-$\beta$ model that we studied in Ref \onlinecite{nikulsin2024an}, the $O(\epsilon)$ correction to the magnetic field is overconstrained, being governed by both a Grad-Shafranov equation and the requirement that flux surfaces exist ($\BB\cdot\nabla\Psi = 0$). This model is significantly more complicated than that of Ref \onlinecite{nikulsin2024an}, and as such, it does not seem to admit purely analytic solutions, except for trivial situations, such as axisymmetry or the large aspect ratio limit. Nevertheless, it enables us to disentangle the overconstraining problem in hybrid quasiaxisymmetric devices, similar to that studied by Henneberg and Plunk in Ref \onlinecite{henneberg2024compact}, by treating the problem in a piecewise manner, considering the axisymmetric and non-axisymmetric regions separately. We then apply standard numerical methods to solve the equations of our model in the two regions. Finally, we use this solution as an initial guess for a stellarator optimization using SIMSOPT\cite{landreman2021simsopt}.

In this paper, we have only considered one particular class of compact stellarators, namely, hybrid quasiaxisymmetric devices that have a purely axisymmetric region, and which allow $\Psi$ to be approximated as a function of just $B_v$ in the non-axisymmetric region, as our method of resolving the overconstraining problem relies on such an approximation. However, there are many more compact quasisymmetric stellarators that do not fit this mold, with NCSX being one prominent example. If we were to attempt to use the present method on a stellarator that does not have an axisymmetric region, we would have to force $\Psi$ to be approximately a function of just $B_v$ everywhere, which is equivalent to $|\BB|$ being approximately a flux function, a situation that can only occur in axisymmetry\cite{schief2003nested}. Furthermore, we can not perform a subsidiary expansion around such an axisymmetric configuration due to it being incompatible with the ordering $|\BB-\nabla\chi|/B_v \ll 1$, which we use to derive the present model\cite{palumbo1968some}. In the future, to overcome this limitation, we intend to implement a numerical solver for the general problem, which will look for a solution in the space of functions of $C_1,C_2$ that minimizes the error in the Grad-Shafranov equation \eqref{eq:GS}.

\section*{Author Declarations}

The authors have no conflicts of interest to disclose.

\begin{acknowledgments}
The authors thank Sophia Henneberg and Gabriel Plunk for providing the VMEC equilibria and vacuum field data for the device that they presented in Ref \onlinecite{henneberg2024compact}, and for fruitful discussions. The authors also thank Per Helander, Elizabeth Paul and Rogerio Jorge for fruitful discussions.

This research was supported by a grant from the Simons Foundation/SFARI (560651, AB) and the Department of Energy Award No. DE-SC0024548. Some computations were performed on the HPC system Viper at the Max Planck Computing and Data Facility (MPCDF).
\end{acknowledgments}

\section*{Data Availability Statement}

The data that support the findings of this study are available from the corresponding author upon reasonable request.

\appendix
\section{The infinitesimal generator of quasisymmetry}\label{sec:infgen}

We can show that the infinitesimal generator of quasisymmetry, $\vec u$, as defined in Ref. \onlinecite{rodriguez2020necessary}, has a simple form when written in terms of the characteristics $C_1,C_2$, namely:
\begin{equation}
    \vec u = \nabla C_1\times\nabla C_2.\label{eq:infgen}
\end{equation}
It is immediately obvious that, up to the order of this model, $\vec u$ satisfies two of the three necessary conditions from Ref. \onlinecite{rodriguez2020necessary}, namely $\nabla\cdot\vec u = 0$ and $\vec u\cdot\nabla|\BB| = 0$, since $|\BB| = B_v + O(\epsilon^2)$.

The third condition is written as follows:
\begin{equation}
    \BB\times\vec u = \nabla\Phi,
\end{equation}
where $\Phi$ is a flux function. Using the equalities \eqref{eq:Bfield} and \eqref{eq:infgen}, we have
\begin{equation}
    \begin{aligned}
        \BB\times\vec u &= [\nabla\chi\cdot\nabla C_2 + (\nabla A\times\nabla\chi)\cdot\nabla C_2]\nabla C_1 \\
        &- [\nabla\chi\cdot\nabla C_1 + (\nabla A\times\nabla\chi)\cdot\nabla C_1]\nabla C_2.\label{eq:Bxu}
    \end{aligned}
\end{equation}
Using \eqref{eq:QSC}, the component in the $\nabla C_1$ direction can be written as
\begin{equation}
    \begin{aligned}
        \nabla\chi\cdot\nabla C_2 &+ \left(\frac{\partial a}{\partial B_v} - \frac{\partial\Psi}{\partial C_1} - \frac{\partial I}{\partial B_v}\right)(\nabla B_v\times\nabla\chi)\cdot\nabla C_2 \\
        &- \frac{\partial I}{\partial s}(\nabla s\times\nabla\chi)\cdot\nabla C_2.
    \end{aligned}
\end{equation}
Further, we can observe that the triple products above can be simplified: $(\nabla B_v\times\nabla\chi)\cdot\nabla C_2 = -g^{-1/2}\partial C_2/\partial s = -1$ and $(\nabla s\times\nabla\chi)\cdot\nabla C_2 = g^{-1/2}\partial C_2/\partial B_v$. Using also the definition of $I$, we have
\begin{equation}
    g^{\chi s}\frac{\partial C_2}{\partial s} + B_v^2\frac{\partial C_2}{\partial\chi} - \frac{\partial a}{\partial B_v} + \frac{\partial\Psi}{\partial C_1} + \frac{\partial I}{\partial B_v}.
\end{equation}
Finally, we can insert equation \eqref{eq:dIdBv} into the above expression, which will result in all terms canceling, except for $\partial\Psi/\partial C_1$. Now consider the component in the $\nabla C_2$ direction in equation \eqref{eq:Bxu}. Using \eqref{eq:QSC} again, it can be written as
\begin{equation}
    -g^{\chi B_v} + \frac{\partial\Psi}{\partial C_2}(\nabla C_2\times\nabla\chi)\cdot\nabla B_v + \frac{\partial I}{\partial s}(\nabla s\times\nabla\chi)\cdot\nabla B_v = \frac{\partial\Psi}{\partial C_2},
\end{equation}
where we have again simplified the triple products, as above, and used the definition of $I$, resulting in the cancellation. Equation \eqref{eq:Bxu} can thus be written as
\begin{equation}
    \BB\times\vec u = \frac{\partial\Psi}{\partial C_1}\nabla C_1 + \frac{\partial\Psi}{\partial C_2}\nabla C_2 = \nabla\Psi.
\end{equation}
Therefore, $\vec u$, as defined in \eqref{eq:infgen}, satisfies all three conditions given in Ref. \onlinecite{rodriguez2020necessary}, and so is an infinitesimal generator of \emph{weak} quasisymmetry\cite{rodriguez2022quasisymmetry}. It is also an infinitesimal generator of \emph{strong} quasisymmetry, since we have imposed magnetohydrostatic force balance in this model, under which \emph{weak} and \emph{strong} quasisymmetry are equivalent.

\bibliography{references}% Produces the bibliography via BibTeX.

\end{document}